\documentclass[conference]{IEEEtran}
\IEEEoverridecommandlockouts
\usepackage{amsmath,amsfonts,amssymb}
\usepackage{graphicx}
\usepackage{cite}
\usepackage{cmap}
\usepackage{tabularx,ragged2e,booktabs}
\usepackage{caption}
\usepackage{subcaption}
\usepackage{tikz}
\usepackage{siunitx}
\usepackage{footnote}
\usepackage{multirow}
\usepackage{url}
\usepackage{hyperref}
\usepackage{algorithm}

\usetikzlibrary{shapes.multipart}
\makesavenoteenv{tabular}
\usepackage[left=0.625in, right=0.625in, top=0.75in, bottom=1.05in]{geometry}
\graphicspath{{pics/}}

\IEEEoverridecommandlockouts
\newcommand\copyrighttext{%
\footnotesize \textcopyright \enspace 2025 IEEE. Personal use of this material is permitted. Permission from IEEE must be obtained for all other uses, in any current or future media, including reprinting/republishing this material for advertising or promotional purposes, creating new collective works, for resale or redistribution to servers or lists, or reuse of any copyrighted component of this work in other works. DOI: \href{https://doi.org/10.1109/BlackSeaCom65655.2025.11193918}{10.1109/BlackSeaCom65655.2025.11193918}
}
\newcommand\copyrightnotice{%
\begin{tikzpicture}[remember picture,overlay]
\node[anchor=south] at (current page.south) {\fbox{\parbox{\dimexpr\textwidth-\fboxsep-\fboxrule\relax}{\copyrighttext}}};
\end{tikzpicture}%
}

\begin{document}

\title{Is It Worth to Use Feedback Channel \\ in 5G V2X Platoon Scenarios?
\thanks{The research was supported by the Ministry of Science and Higher Education of the Russian Federation, project № FFNU-2025-0030.}
}

\author{\IEEEauthorblockN{Dmitry Bankov, Artem Krasilov, Artem Otmakhov, Pavel Savlukovich, Evgeny Khorov} 
	
	\IEEEauthorblockN{Institute for Information Transmission Problems of the Russian Academy of Sciences, Moscow, Russia}
	
	\IEEEauthorblockN{Email: \{bankov, krasilov, otmakhov, savlukovich, khorov\}@wnlab.ru}
}
\maketitle
\copyrightnotice

\begin{abstract}
	5G Vehicle-to-Everything (V2X) is a new technology developed by 3GPP to support inter-vehicle communication. In contrast to 4G V2X which allows only broadcast communication, 5G V2X enables groupcast and unicast communication. Such types of communication are needed for new V2X scenarios: platooning, extended sensors, remote driving, etc. To improve the data transmission reliability and assist in the selection of the transmission parameters in these scenarios, 5G V2X introduces a feedback channel that allows receivers to send acknowledgments in response to data packets. However, some part of the overall resource shall be allocated for the feedback channel, which reduces the amount of channel resources available for data transmission. In this paper, we consider a scenario with a platoon, which generates groupcast traffic, and surrounding vehicles, which generate legacy broadcast traffic. Using extensive simulations in NS-3, we analyze how the usage of the feedback channel influences the overall system capacity. Our results show that depending on the platoon size, groupcast, and broadcast traffic intensities, and their quality of service requirements, the usage of the feedback channel can in some cases significantly increase the system capacity (up to 2x), while in other cases it almost halves the system capacity. We explain the reasons for such effects and discuss how to adaptively select the feedback channel parameters.   
\end{abstract}

\begin{IEEEkeywords}
	5G, V2X, IoV, Mode 2, DENM, acknowledgements, vehicle platooning
\end{IEEEkeywords}

\section{Introduction}
\label{sec:intro}
Nowadays, the paradigm of autonomous vehicles receives much attention both in academia and industry.
There are several approaches for managing such vehicles. One of the popular approaches considered by many vehicle manufacturers is to use local on-board sensors and autonomous control algorithms~\cite{parekh2022review}. However, with this approach, a vehicle cannot sense vehicles hidden by the obstacles (e.g., buildings at the crossroad).  Thus, another approach is to supplement information from local sensors with the additional information provided by surrounding vehicles and roadside infrastructure delivered via wireless networks. Such a communication paradigm is called the Internet of Vehicles (IoV)~\cite{ji2020survey}.
One of the key problems for IoV is to enable communication between vehicles and infrastructure with low latency and high reliability.
To solve this problem, the 3rd Generation Partnership Project (3GPP) has recently introduced the 5G Vehicle-to-Everything (V2X) technology~\cite{garcia2021tutorial}. 5G V2X provides the specifications to enable reliable and low-latency communication between vehicles and other surrounding objects in different scenarios (e.g., autonomous driving, vehicle platooning, etc).
The specific requirements for the Quality of Service (QoS) can be expressed in terms of latency budget $D^{QoS}$ and maximal packet loss rate $PLR^{QoS}$, which depend on the required Level of Automation (LoA) and are presented in the 3GPP specifications~\cite{3gpp-v2x-scenarios}. For example, high LoA scenarios (e.g., platooning) require $D^{QoS}$ of the order of ${10}$~ms and $PLR^{QoS}$ of the order of $1\%$.

To provide direct communication between vehicles, 5G V2X introduces a sidelink interface in addition to the basic for cellular systems uplink/downlink interfaces~\cite{bazzi2021design}. At the MAC layer, vehicles can use either a centralized channel access method called Mode 1 (i.e., all transmissions are controlled by the base station) or a distributed channel access method called Mode 2 (i.e., each vehicle autonomously selects resources for its transmission). As shown in~\cite{bankov2023analytical, bankov2024enhancing}, Mode 2 is the only channel access method that can be used in scenarios with high LoA and strict latency requirements because Mode 1 requires long control information exchange with the base station.

In 5G V2X networks, vehicles (hereinafter called UEs) generate heterogeneous traffic that consists of two types of messages. First, UEs can periodically broadcast Cooperative Awareness Messages (CAMs), which contain information about their current position, velocity, acceleration, etc. Second, when a UE detects a critical situation on the road at a random time moment, it broadcasts a Decentralized Environmental Notification Message (DENM) to alert the neighboring UEs. DENMs shall be delivered with low latency and high reliability because their loss can lead to dangerous consequences.
In this paper, we assume that non-overlapping frequency resources are allocated for transmitting CAMs and DENMs, which allows avoiding interference between different types of traffic and increasing transmission reliability~\cite{yin2022design}.
Thus, in this paper, we only consider sporadic DENM traffic.

In contrast to the previous generation 4G V2X technology which allows only broadcast communication, 5G V2X enables groupcast and unicast communication. Such types of communication are needed for new V2X scenarios: platooning, extended sensors, remote driving, etc~\cite{3gpp-v2x-study}. In the paper, we focus on the platoon scenario since it requires communication between a group of vehicles moving together. However, the results are applicable to other scenarios which require unicast and groupcast communication.

To increase the reliability of data transmission in the case of unicast and groupcast communication, 5G V2X introduces the feedback channel. When a UE receives a data packet, it can notify the sender about the success or failure of decoding a packet (i.e., send an acknowledgment) via a feedback channel. On the one hand, the usage of the feedback channel allows avoiding unnecessary blind retransmissions if all receivers have acknowledged the packet reception. On the other hand, the feedback channel reduces the amount of channel resources available for data transmissions. In this paper, we study in detail how the usage of the feedback channel (i.e., the acknowledgments) influences the overall system capacity.  We consider a scenario with a platoon that generates groupcast traffic and surrounding vehicles which generate legacy broadcast traffic. Based on extensive simulations in NS-3, we show that depending on the scenario (i.e., the platoon size, groupcast and broadcast traffic intensities, and their QoS requirements), the usage of the feedback channel can, in some cases, significantly increase the system capacity, while in other cases drop the capacity. We explain the reasons for such effects and discuss how to adaptively select the feedback channel parameters.


The rest of the paper is organized as follows.
Section~\ref{sec:psfch} provides an overview of 5G V2X PHY and MAC layers, including the description of the feedback channel.
In Section~\ref{sec:works}, we review the related works.
Section~\ref{sec:scenario-problem} describes the considered scenario and the problem statement.
Performance evaluation of the feedback channel efficiency is presented in Section~\ref{sec:performance}.
We conclude the paper in Section~\ref{sec:conclusion}.

\section{5G V2X PHY\&MAC Overview}
\label{sec:psfch}

In this section, we shortly describe the 5G V2X focusing on PHY and MAC layers. The PHY layer operates as follows. In the time domain, the channel resources are divided into slots each consisting of 14 Orthogonal Frequency-Division Multiplexing (OFDM) symbols. In the frequency domain, the channel is divided into several subchannels of equal width. When a UE has a pending data packet, it selects a slot and a continuous set of subchannels for transmission of this packet.         

\begin{figure}[!t]
	\centering
	\begin{subfigure}{.5\textwidth}
		\centering
		\includegraphics[width=0.9\linewidth]{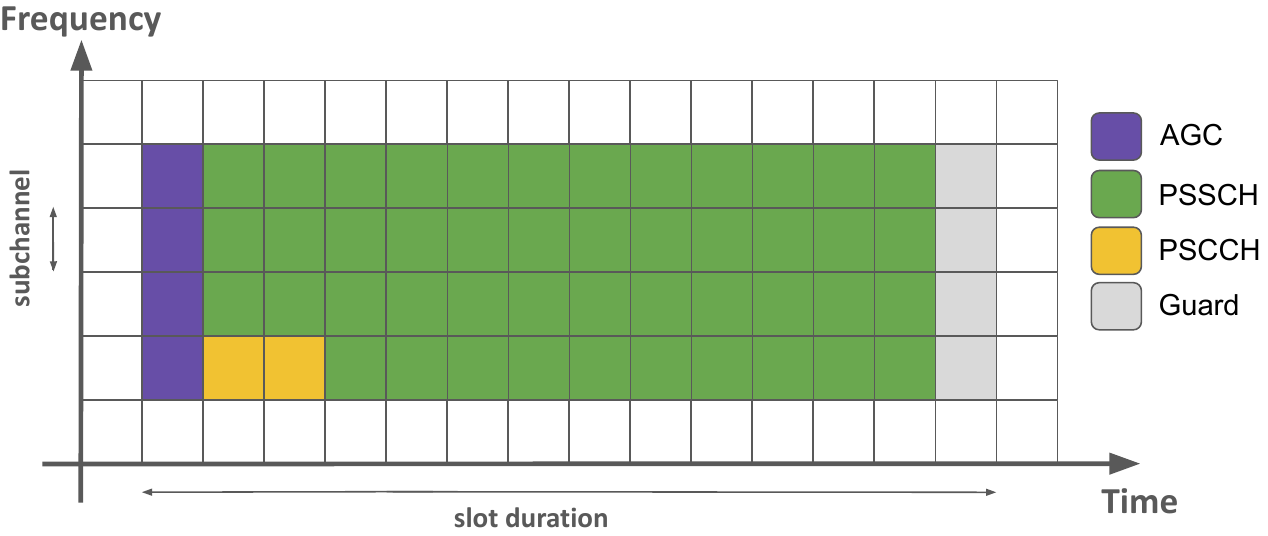}
		\caption{without PSFCH}
		\label{fig:ack0}
	\end{subfigure}%

	\begin{subfigure}{.5\textwidth}
		\centering
		\includegraphics[width=0.9\linewidth]{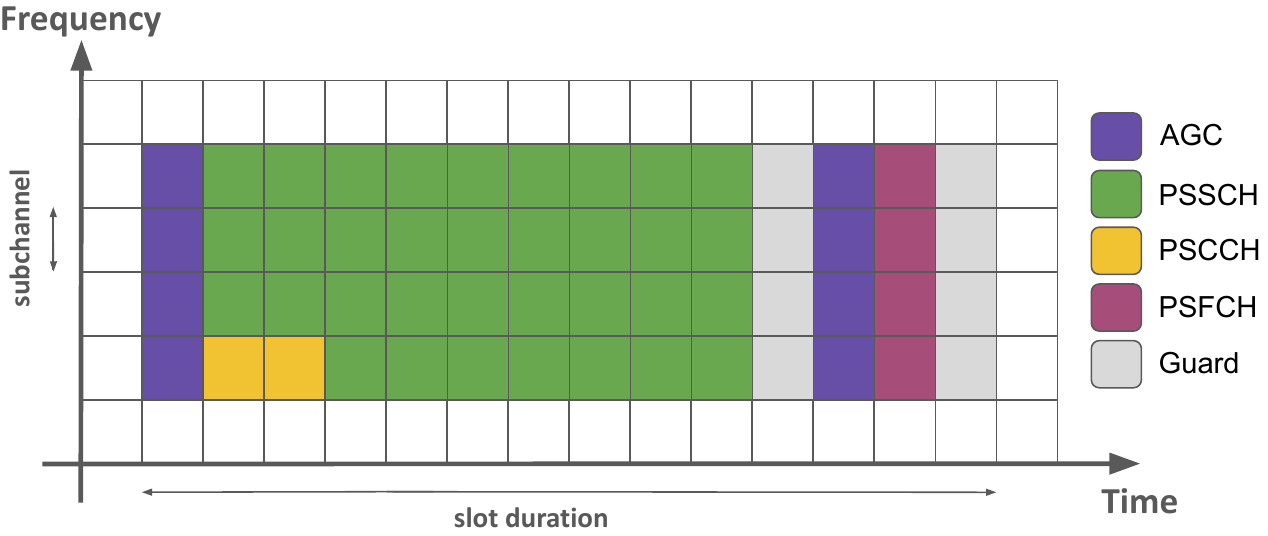}
		\caption{with PSFCH}
		\label{fig:ack1}
	\end{subfigure}

	\caption{Slot structure.}
	\label{fig:slot}
\end{figure}

Each slot consists of (see Fig.~\ref{fig:ack0})~\cite{3gpp-phy-channels}:
(i) Automatic Gain Control (AGC) symbol which is used for adjusting the receiver amplifier gain,
(ii) the Physical Sidelink Control Channel (PSCCH), which carries control information, such as Modulation and Coding Scheme (MCS) used for data,
(iii) the Physical Sidelink Shared Channel (PSSCH), which carries user data,
(iv) Guard symbol used to avoid the consequent slots signals overlapping.
PSCCH is encoded with the most robust MCS, while the MCS for PSSCH can be adaptively selected based on the current channel conditions and traffic. The receiving UE first decodes PSCCH. Only if PSCCH is decoded successfully, the UE can decode the user data in PSSCH.

For scenarios with unicast and groupcast traffic, 5G V2X specifications allow configuring resources for the Physical Sidelink Feedback Channel (PSFCH). PSFCH occupies one or several OFDM symbols which are used for the transmission of acknowledgments (ACK) or negative acknowledgments (NACK) in response to data packets. Since PSFCH is transmitted by the receiving UE, it is accompanied by the AGC and Guard symbols (see Fig.~\ref{fig:ack1}).  ACK/NACKs of different UEs are multiplexed in frequency and code domains ~\cite{garcia2021tutorial}.  Note that PSFCH is configured for the whole network: all UEs, including those that generate broadcast traffic and do not require acknowledgments, cannot transmit data in OFDM symbols allocated for PSFCH.

Figure~\ref{fig:slot} shows the slot configurations with and without PSFCH considered in this paper.
In the absence of PSFCH, 12 out of 14 OFDM symbols can be used for data transmission (i.e., for PSSCH).
When PSFCH is configured, one OFDM symbol is allocated for the PSFCH channel, one symbol for AGC, and one symbol for Guard after  PSFCH.
Thus, the usage of the feedback channel reduces the number of OFDM symbols available for user data from twelve down to nine, which results in 25\% loss of channel resources for PSSCH.

MAC layer is responsible for: (i) selecting channel resources for PSCCH/PSSCH transmission, (ii) selecting transmission parameters, such as MCS and the number of transmission attempts~\cite{3gpp-mac}.
In this paper, we assume that UEs use the distributed channel access method Mode 2. With Mode 2, each UE senses the channel and determines the set of slots and subchannels reserved by other UEs. The information about the reserved resources is provided in PSCCH. For example, a UE can use PSCCH to reserve several slots and subchannels for its next (re)transmissions. When a UE selects a resource for its own transmission, it randomly selects a slot and a continuous set of subchannels that are not reserved by other UEs. The number of subchannels required for a packet transmission depends on the packet size, the subchannel width, and the used MCS. 

To improve reliability, each packet is transmitted several times. When PSFCH is not configured, UE has no feedback from the receiving UEs, thus, it repeats the transmission of a packet $K$ times, where $K$ is the parameter selected by the MAC layer. In particular, $K$ should be lower than the packet latency budget divided by the slot duration. When PSFCH is configured, the UE can stop the transmission of a packet when all receiving UEs report ACK. In this case, $K$ is the maximum number of transmission attempts. For each packet, the real number of transmission attempts will be lower or equal to $K$. Thus, the usage of PSFCH can reduce the amount of resources used by UEs for retransmissions and reduce interference between UEs.

To sum up, the usage of PSFCH can increase reliability and reduce the number of transmission attempts by canceling the pending retransmissions if the PSSCH has been successfully decoded by all receivers.
However, PSFCH occupies channel resources that could be used for PSSCH. In this paper, we study in detail whether the loss of resources for PSFCH can be compensated by more efficient usage of channel resource thanks to ACKs/NACKs feedback.


\section{Related Works}
\label{sec:works}
5G V2X Mode 2 with a feedback channel has been studied in several papers.
In particular, papers~\cite{garcia2021tutorial,lien20203gpp,ali20213gpp} describe how the feedback channel works.
The authors claim that PSFCH allows making feedback-based retransmissions, and this method is resource-efficient, but increases the latency.
However, the authors do not provide numerical results regarding the efficiency of the feedback channel.

Only a few works study the efficiency of the feedback channel.
For example, authors of \cite{liu2023towards} introduce blind and feedback-based retransmission strategies, as well as the slot structure for both cases.
The results show that blind retransmissions are better in terms of latency and communication range.
However, only a simple point-to-point scenario with two UEs is considered, which does not correspond to realistic scenarios recommended by 3GPP.

Paper~\cite{kim2019efficient} considers a scenario in which a platoon is surrounded by non-platoon UEs sending unicast packets. Simulation results demonstrate that blind retransmissions achieve higher reliability, while the feedback mechanism increases spectral efficiency. However, authors only consider periodic traffic with basic QoS requirements (i.e., reliability of the order of 90\%), while this paper considers new V2X scenarios with high LoA and strict QoS requirements.

Authors of~\cite{kuo2021reliable} state that the feedback channel is effective in scenarios with moderate latency requirements, while blind retransmissions shall be used in case of strict latency requirements.
However, the paper does not provide any numerical results which prove this statement and does not provide a formal rule to select the retransmission strategy depending on latency requirement.

Therefore, it is necessary to consider different latency requirements and analyze in which scenarios the usage of the feedback channel increases the network capacity.
Our paper fills this gap.
Moreover, in contrast to existing papers, we consider scenarios with sporadic traffic and strict QoS requirements inherent to future V2X scenarios.

\section{Scenario and Problem Statement}
\label{sec:scenario-problem}
We consider a 5G V2X network consisting of $N$ UEs moving at a fixed speed on a highway. Following the methodology used in related works~\cite{kim2019efficient,bankov2023analytical}, the distance between neighboring UEs has an exponential distribution with the parameter $\phi$. The path loss in dB at a distance $d$ is given by the log-distance propagation model: $L(d)=L_0+10 \cdot n\cdot\log_{10}(\frac{d}{d_0})$.

\begin{figure}[!t]
	\centering
	\includegraphics[width=1.0\linewidth]{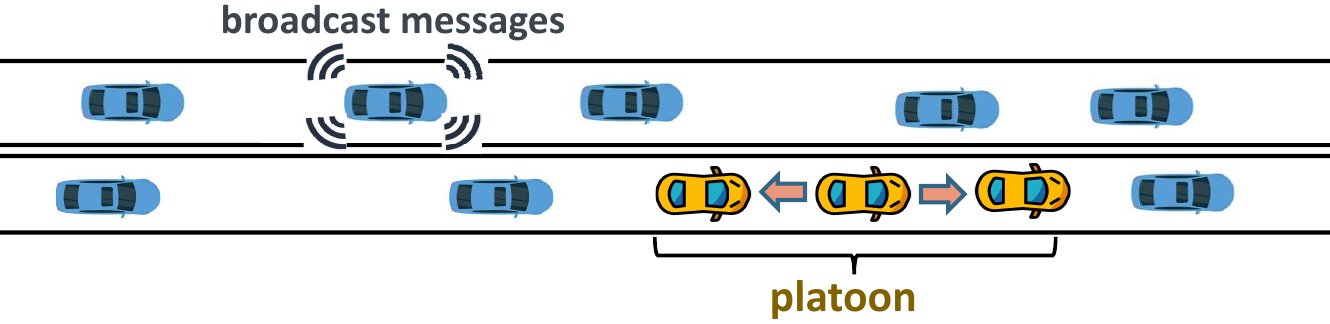}
	\caption{The highway scenario.}
	\label{fig:scenario}
\end{figure}

As illustrated in Fig.~\ref{fig:scenario}, $N_g$ UEs form a platoon. Inside the platoon, each UE generates a Poisson flow of groupcast DENM packets with intensity $\lambda_g$ packets per second. Packets generated by any platoon member are addressed to all platoon members. The remaining $N_b$ UEs on a highway generate broadcast DENM packets. Each UE generates packets with intensity $\lambda_b$. We estimate the reliability of delivering broadcast packets by considering only receivers located at a distance not exceeding the communication range $R$. The specific requirements on $R$ for various scenarios are given in \cite{3gpp-v2x-scenarios}.

The channel consists of $B$ subchannels, and the slot duration is $\tau$. We assume that all UEs transmit $L$-byte DENM packets using  Mode 2. All UEs use MCS $\chi$ for PSSCH and transmit the signals with the same power $S$ according to~\cite{romeo2021supporting,bankov2024enhancing}.
The maximum number of attempts to transmit a single packet is $K_g$ for the platoon members and $K_b$ for the remaining UEs generating broadcast packets. Note that when PSFCH is configured, the real number of transmission attempts for platoon members is lower than or equal to $K_g$. Otherwise, if  PSFCH is not configured, each platoon member repeats the transmission of each packet exactly $K_g$ times.


We consider two configurations of the time slot.
\begin{enumerate}
	\item Without PSFCH: twelve symbols are allocated for PSSCH (see Fig.~\ref{fig:ack0}), all UEs make blind retransmissions.
	\item With PSFCH: only nine symbols can be used for PSSCH (see Fig.~\ref{fig:ack1}), platoon members send ACKs/NACKs using PSFCH to control the number of retransmissions, while the remaining UEs make blind retransmissions.
\end{enumerate}
We note that, in the second configuration, reducing the number of symbols allocated to PSSCH increases the number of subchannels required for data packet transmission.
Moreover, in configuration with PSFCH, the sender waits for ACKs from receivers, which are generated with delay $D^{ACK}$ (of the order of several slots).  Thus, the maximum number of transmission attempts for groupcast packet $K_g$ can be lower than the maximum number of repetitions $K_b$ of a broadcast packet.

In our study, we consider the following network performance indicators.
For non-platoon UE $i$ the packet loss ratio $PLR^b_i$ is calculated as the average fraction of receivers in a circle of radius $R$ which do not decode the broadcast packet for any transmission attempt.
For the entire network, the average packet loss ratio $PLR^b$ is estimated as the average value of $PLR^b_i$ over all UEs.
For platoon UE $j$, $PLR^g_j$ is the fraction of groupcast packets not delivered to at least one platoon member after all transmission attempts.
The average packet loss ratio $PLR^g$ is estimated as the average value of $PLR^g_j$ for all platoon members.
In the paper, we fix the load $\lambda_b$ for the broadcast traffic and determine the maximum load that can be supported by the network for the groupcast traffic (i.e., the traffic generated within the platoon). We define the network capacity for groupcast traffic as follows:
\begin{equation*}
\begin{split}
	& C_g(\lambda_b) = \max \{\lambda_g: PLR^b(\lambda_b, \lambda_g)\leq PLR^{QoS}, \\
	& PLR^g(\lambda_b, \lambda_g) \leq PLR^{QoS} \} 
\end{split}
\end{equation*}
where  $PLR^{QoS}$ is the PLR requirement for both broadcast and groupcast traffic.

In the paper, we study in detail how $C_g$ depends on the feedback channel configuration and other scenario parameters (i.e., platoon size, broadcast traffic intensity, QoS requirements, etc) and provide recommendations in which cases the usage of the feedback channel is fruitful for increasing network capacity.

\section{Performance Evaluation}
\label{sec:performance}
\subsection{Simulation setup}
\label{subsec:simulation}
To study the efficiency of PSFCH, we use the NS-3 network simulator~\cite{ns3} with the V2X model developed in~\cite{cttc}. We extended this model by adding the scenario described in Section~\ref{sec:scenario-problem} and implementing the PSFCH model.
The main simulation parameters are listed in Table~\ref{tab:parameters}.

\begin{table}[!t]
	\caption{Main simulation parameters.}
	\label{tab:parameters}
	\begin{center}
		\begin{tabular}{|l|c|}
			\hline
			\textbf{Parameter} & \textbf{Value} \\
			\hline
			Number of UEs $N$ & 200 \\
			Transmission power $S$ & 23 dBm \\
			Noise figure & 5 dB \\
			MCS for PSSCH $\chi$ & 6  \\
			Error model & EESM~\cite{lagen2020new} \\
			Slot duration $\tau$ & $\SI{500}{\us}$ \\
			Number of subchannels $B$ & 10 \\
			Average distance between UEs $\phi^{-1}$ & 10 m \\
			Packet size $L$ & 290 bytes \\
			Reference distance $d_0$ & 1 m \\
			Reference loss at reference distance $L_0$ & 46,7 dB\\
			Path loss exponent $n$ & 3 \\
			Communication range $R$ & 200 m \\
			Maximum packet loss rate $PLR^{QoS}$ & $10^{-2}$ \\
			Acknowledgement delay $D^{ACK}$ & $\SI{2}{\ms}$ \\
			\hline
		\end{tabular}
	\end{center}
\end{table}

We study the efficiency of PSFCH in the presence of both broadcast and groupcast traffic.
Each scenario considered below is determined by broadcast traffic load $\lambda_b$, the number of platoon members $N_g$, and latency budget  $D^{QoS}$.
Unless otherwise stated, we consider the following default scenario parameters: $\lambda_b = 1.8$ packets/s, $N_g = 5$, and $D^{QoS} = 10$ ms.

\subsection{Impact of the broadcast traffic intensity}
\label{subsec:lambda_b}
\begin{figure}[!t]
	\centering
	\begin{subfigure}{.5\textwidth}
		\centering
		\includegraphics[width=0.9\linewidth]{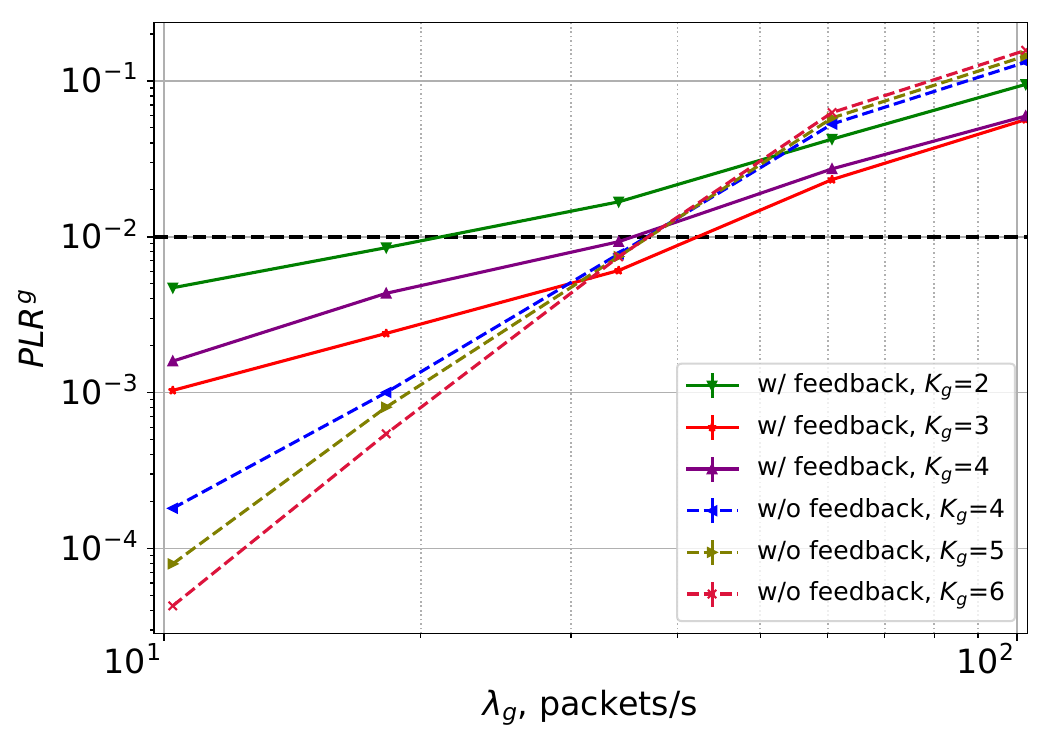}
		\caption{for platoon UEs}
		\label{fig:rate_plr_g}
	\end{subfigure}%

	\begin{subfigure}{.5\textwidth}
		\centering
		\includegraphics[width=0.9\linewidth]{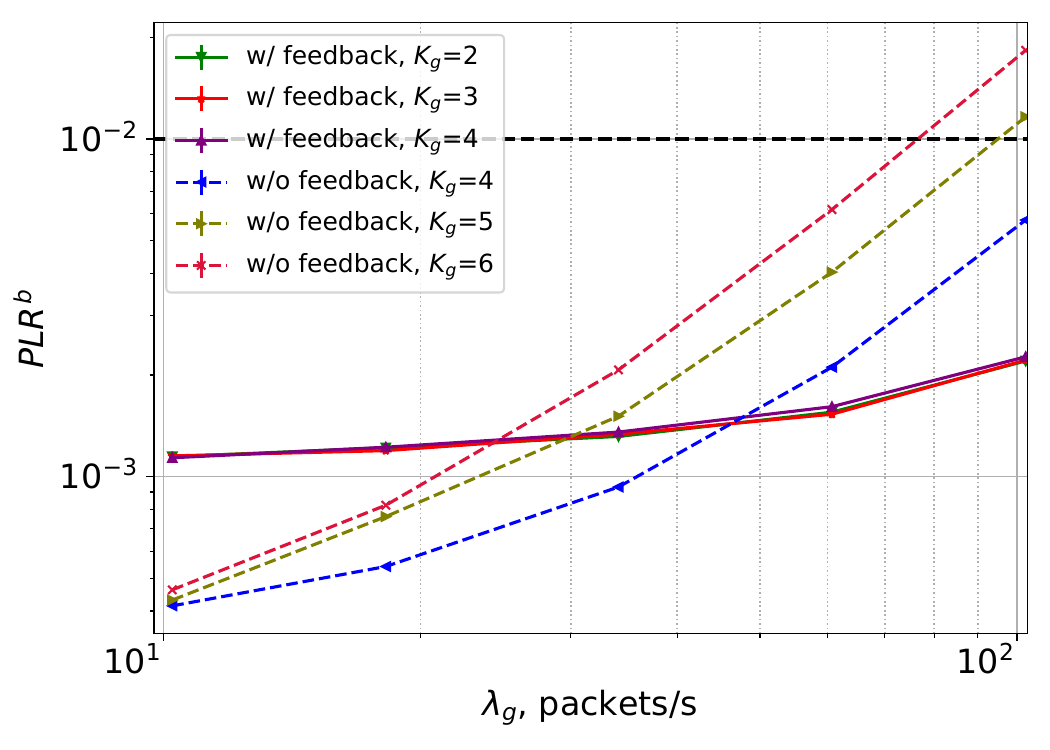}
		\caption{for non-platoon UEs}
		\label{fig:rate_plr_b}
	\end{subfigure}
	
	\caption{PLR as the function of $\lambda_g$.}
	\label{fig:rate_plr}
\end{figure}

Let us consider a scenario with the default parameters and analyze how $PLR^b$ and $PLR^g$ depend on the traffic intensity $\lambda_g$ of platoon members (see Fig.~\ref{fig:rate_plr}).
Based on the definition provided in Section~\ref{sec:scenario-problem}, we find the network capacity as follows.
We determine intensities of groupcast traffic for which  $PLR^b$ and $PLR^g$ curves intersect $PLR^{QoS}$ limit (see the dotted black horizontal line). The network capacity is the minimum of these values.
We can see that the network capacity depends on: (i) whether we use PSFCH or not, and (ii) the number of transmission attempts for groupcast packets $K_g$. Using exhaustive search, we can find the optimal $K_g$ which maximizes the network capacity. In particular, for the considered scenario, without PSFCH, the optimal $K_g=5$, while, with PSFCH, the optimal is $K_g=3$.


\begin{figure}[!t]
	\centering
	\includegraphics[width=1.0\linewidth]{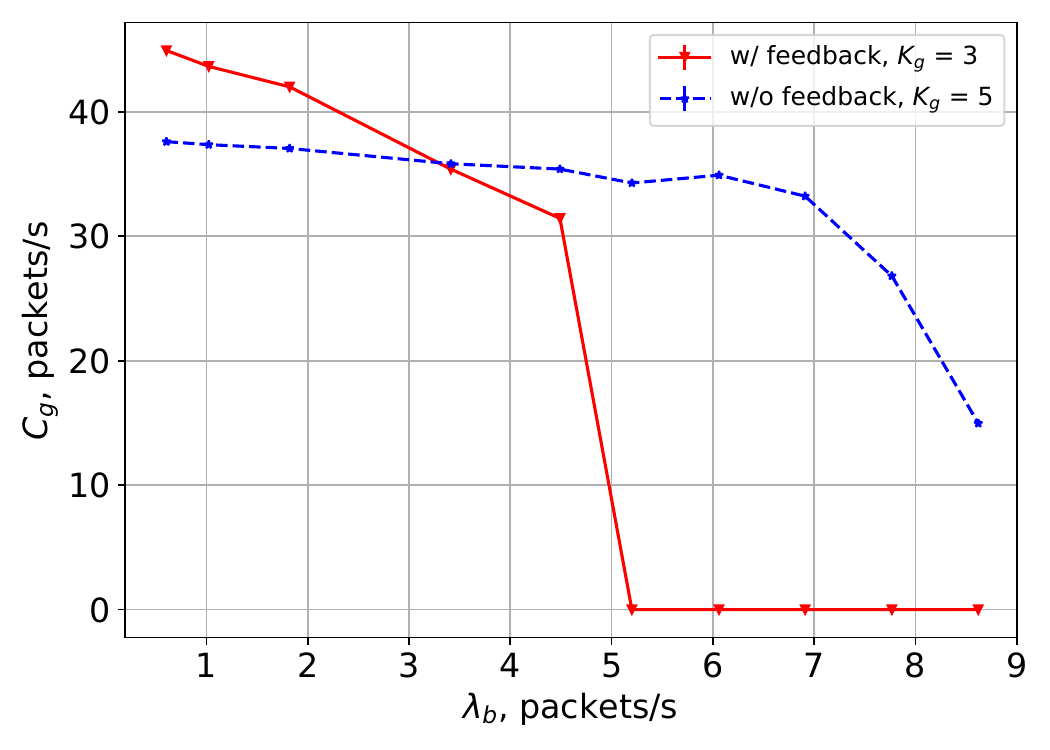}
	\caption{The network capacity as the function of broadcast traffic intensity $\lambda_b$.}
	\label{fig:rate_capacity}
\end{figure}

Let us further study how the broadcast traffic intensity $\lambda_b$ influences the network capacity.
For that, we performed the procedure described above for various $\lambda_b$ values.
Fig.~\ref{fig:rate_capacity} shows the results for two configurations: with and without PSFCH, each with the optimal number of attempts $K_g$.
First, we see that the usage of PSFCH increases the network capacity for low intensity of broadcast traffic ($\lambda_b < 3.4$).
Note that $\lambda_b = 3.4$ corresponds to a 10\% occupation of channel resources by the broadcast traffic.
For example, with $\lambda_b = 0.6$, the usage of PSFCH increases the network capacity by 20\%.
Second, we see a sharp degradation of the network capacity for the case when PSFCH is used and $\lambda_b$ reaches $5$ packets/s.
A zero network capacity means that the $PLR^{QoS}$ requirement cannot be satisfied for any $\lambda_g$.

Without PSFCH network capacity degrades for higher values when $\lambda_b=8...9$. The observed effects are explained by the following facts. First, when  PSFCH is not configured, the PSSCH channel has 25\% more available channel resources. Thus, the probability of packet collisions reduces and the network performance degradation is observed at higher values of $\lambda_b$. Second, when $\lambda_b$ is low, the probability of packet collisions becomes also low. Thus, the usage of PSFCH allows effectively reducing the number of retransmissions for platoon UEs compared to blind retransmissions. So, we can conclude that the usage of PSFCH is fruitful only at a low intensity of broadcast traffic.



\subsection{Impact of the number of platoon members}
\label{subsec:N_g}
\begin{figure}[!t]
	\centering
	\includegraphics[width=1.0\linewidth]{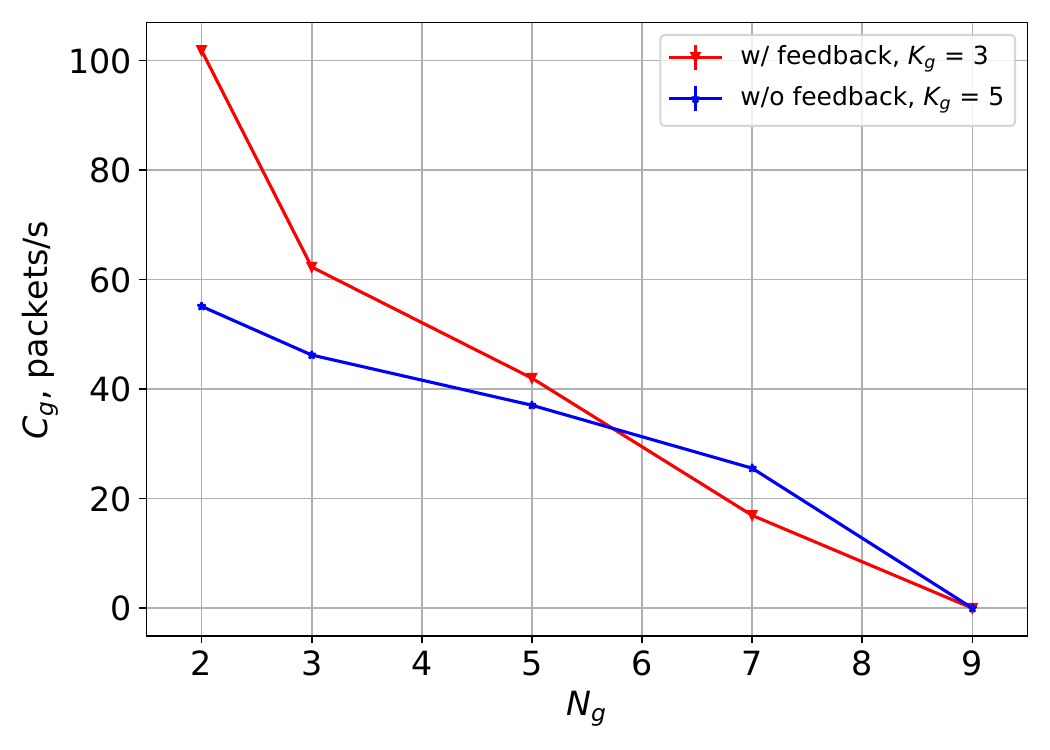}
	\caption{The network capacity as the function of platoon size $N_g$.}
	\label{fig:num_platoons_capacity}
\end{figure}

Figure~\ref{fig:num_platoons_capacity} shows the dependence of the network capacity on the number of platoon members $N_g$.
We can see that significant increase of network capacity with PSFCH is achieved for small platoon sizes (85\% gain at $N_g = 2$).
Further, as the platoon size increases, the efficiency of the PSFCH-based retransmissions significantly decreases, and, at $N_g \geq 6$, its capacity becomes less than in the absence of PSFCH (i.e., for blind retransmissions).
The reason is as follows. For small platoon size, each packet has fewer receivers and the average distance to the furthest receiver becomes smaller. This leads to a higher probability of successful packet delivery to all platoon UEs, which, in turn, increases the efficiency of the feedback mechanism by canceling unnecessary retransmissions.
Thus, the usage of PSFCH provides notable gain only for small platoon sizes.

\subsection{Impact of the latency budget}
\label{subsec:D^QoS}
\begin{figure}[!t]
	\centering
	\includegraphics[width=1.0\linewidth]{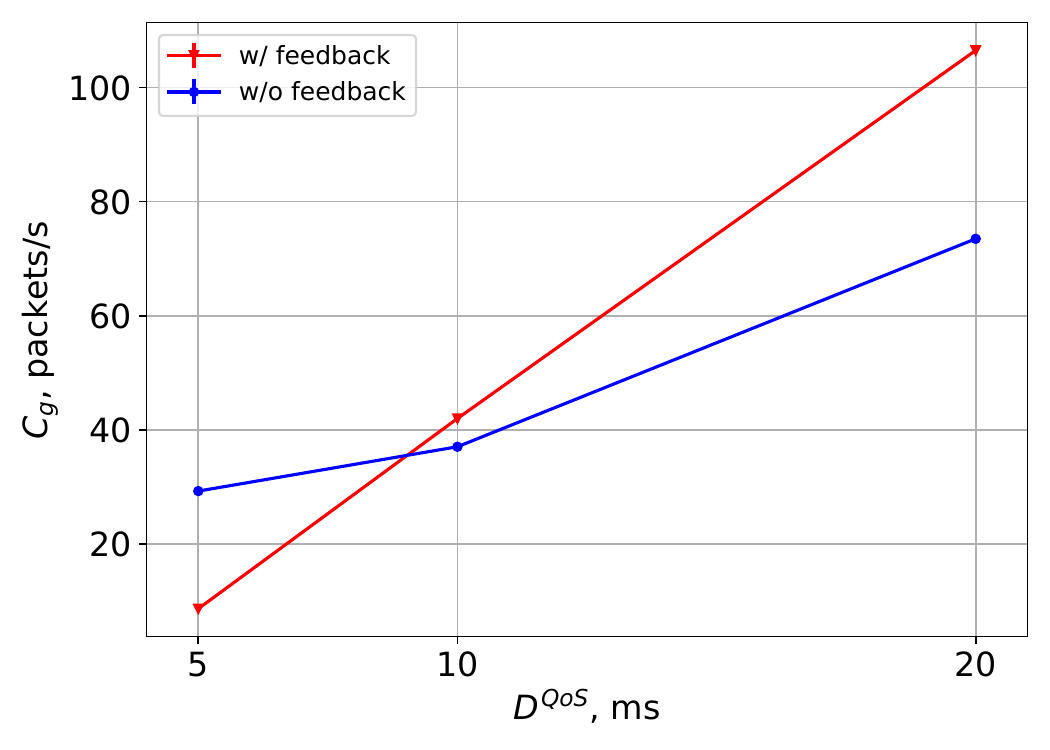}
	\caption{The network capacity as the function of latency budget $D^{QoS}$.}
	\label{fig:d_capacity}
\end{figure}

Let us study the impact of the latency budget $D^{QoS}$ on the network capacity with and without PSFCH.
Figure~\ref{fig:d_capacity} shows that with a higher value of $D^{QoS}=20$ ms, there is a significant increase in the network capacity (up to 45\%) thanks to the feedback channel. In contrast, with a strict latency requirement $D^{QoS}=5$ ms (which is expected to be relevant for 6G networks), the usage of PSFCH  notably reduces capacity (down to 70\%).
The observed effect is explained by the fact that in the case of PSFCH usage, the time between consecutive transmission attempts shall be at least $D^{ACK}$ ($D^{ACK}=2$~ms in our model), while blind retransmissions can be done every slot.
This fact leads to the following consequences.
First, strict $D^{QoS}$ significantly limits the maximum number of transmission attempts $K_g$.
For example, for $D^{QoS}=5$~ms, the maximum number of transmissions with PSFCH is two, while without PSFCH it can be set up to nine.
Thus, for strict $D^{QoS}$, the usage of PSFCH does not allow to provide a high number of transmission attempts required for satisfying high-reliability constraints. 
Second, with high $D^{QoS}$, UEs have enough channel resources for making all transmission attempts needed to provide given reliability, while the feedback channel allows canceling unnecessary retransmissions for each particular packet. 
To sum up, the usage of PSFCH is effective in cases when latency budget $D^{QoS}$ is much higher than $D^{ACK}$.

\section{Conclusion}
\label{sec:conclusion}
In this paper, we considered the 5G V2X platoon scenario with the coexistence of broadcast and groupcast traffic. With simulations, we analyzed in detail how the usage of the feedback channel (PSFCH) influences the network performance depending on the scenario parameters, such as broadcast traffic intensity, platoon size, latency budget, etc. Based on the obtained results, we determined the set of scenarios in which the usage of PSFCH can significantly increase the network capacity. First, the PSFCH should be used in case of low-intensity background broadcast traffic generated by non-platoon UEs (i.e., the broadcast traffic should occupy a small fraction of channel resources, $\sim10\%$). Second, the platoon size should be small to provide a high packet delivery probability for the furthest receiver using the selected MCS. Third,  PSFCH can significantly increase the network capacity when the latency budget is much higher than the time needed for the receiver to generate feedback (i.e., higher than ACK delay). 


We want to emphasize that the overall network performance significantly depends on the considered scenario: the usage of PSFCH can either significantly increase or decrease the network capacity. 
In our future works, we are going to develop a method that will adaptively select PSFCH parameters (i.e., whether to use PSFCH and the period of PSFCH). For that, we are going to extend our analytical model~\cite{bankov2023analytical} developed for the case of blind retransmissions. The model will allow estimating the capacity as the function of scenario parameters and PSFCH parameters. Another direction for future research is to consider the problem of adaptive MCS selection based on the information obtained via the feedback channel.

\bibliographystyle{IEEEtran}
\bibliography{biblio}

\end{document}